\documentclass[twocolumn,apj]{openjournal}

\usepackage[breaklinks,colorlinks,citecolor=blue,urlcolor=blue]{hyperref}
\usepackage{amsmath}

\usepackage[T1]{fontenc}
\usepackage{newtxtext,newtxmath}
\usepackage{graphicx}
\usepackage{amsfonts,ulem}
\usepackage{soul}
\usepackage{hyperref}
\usepackage{enumitem}
\usepackage[hyphens]{xurl}
\usepackage{multirow}

\newcommand{\orcidauthor}[3]{\author{\href{http://orcid.org/#1}{#2$^{#3}$}}}

\renewcommand{\arraystretch}{1.2}  

\shorttitle{Are Lens Models Converging or Diverging?}
\shortauthors{Perera et al.}

\begin{document}

\title{\vspace{-0.8cm}Are Models of Strong Gravitational Lensing by Clusters Converging or Diverging?\vspace{-1.5cm}}

\orcidauthor{0000-0002-4693-0700}{Derek Perera}{1,^\dagger}
\orcidauthor{0000-0002-2901-5260}{John H. Miller Jr.}{1} 
\orcidauthor{0000-0002-6039-8706}{Liliya L. R. Williams}{1}
\orcidauthor{0000-0002-3648-8031}{Jori Liesenborgs}{2}
\orcidauthor{0000-0002-9800-9868}{Allison Keen}{1} 
\orcidauthor{0000-0002-4490-7304}{Sung Kei Li}{3}
\orcidauthor{0000-0001-6636-4999}{Marceau Limousin}{4}

\affiliation{$^1$School of Physics and Astronomy, University of Minnesota, Minneapolis, MN, 55455, USA}
\affiliation{$^{2}$UHasselt – Flanders Make, Expertisecentrum voor Digitale Media, Wetenschapspark 2, B-3590, Diepenbeek, Belgium.}
\affiliation{$^{3}$Department of Physics, The University of Hong Kong, Pokfulam Road, Hong Kong.}
\affiliation{$^{4}$Aix Marseille Univ, CNRS, CNES, LAM, Marseille, France.}

\thanks{\vspace{0.1cm}$^\dagger$Corresponding author: \href{mailto:perer030@umn.edu}{perer030@umn.edu}}

\begin{abstract}\vspace{0.2cm}
    The increasingly large numbers of multiple images in cluster-scale gravitational lenses have allowed for tighter constraints on the mass distributions of these systems. Most lens models have progressed alongside this increase in image number. The general assumption is that these improvements would result in lens models converging to a common solution, suggesting that models are approaching the true mass distribution. To test whether or not this is occurring, we examine a sample of lens models of MACS J0416.1$-$2403 containing varying number of images as input. Splitting the sample into two bins (those including $<150$ and $>150$ images), we quantify the similarity of models in each bin using three comparison metrics, two of which are novel: Median Percent Difference, Frechet Distance, and Wasserstein Distance. In addition to quantifying similarity, the Frechet distance metric seems to also be an indicator of the mass sheet degeneracy. Each metric indicates that models with a greater number of input images are no more similar between one another than models with fewer input images. This suggests that lens models are neither converging nor diverging to a common solution for this system, regardless of method. With this result, we suggest that future models more carefully investigate lensing degeneracies and anomalous mass clumps (mass features significantly displaced from baryonic counterparts) to rigorously evaluate their model's validity. We also recommend further study into alternative, underutilized lens model priors (e.g. flux ratios) as an additional input constraint to image positions in hopes of breaking existing degeneracies. 
\end{abstract}

\section{Introduction}

Present in the literature of cluster scale gravitational lensing for a long time has been the notion that future lens data will reach increasingly larger numbers of lensed multiple images\footnote{Throughout this paper we use ``images" to mean image positions unless otherwise noted.}, resulting in more accurate lens models \citep[e.g.,][]{natarajan24}. With the advent of the James Webb Space Telescope (JWST) these new large datasets are becoming possible, with observations able to attain deeper depths and uncover previously invisible lensed images. This is to say, that as the number of observed images (i.e. lensing constraints) increases for a given cluster lens, our lens modelling reconstructions, regardless of method, will begin to converge towards a common solution.

It should be noted that one will never know the true mass distribution of any cluster. Therefore, the best indicator of an accurate mass reconstruction is the confluence of different reconstructions from different methods. In this work, we assume that if mass reconstructions are converging to a common solution, then these solutions are approaching the true mass distribution. The expectation that very large numbers of images will yield mass maps that are close to true ones is not unfounded and was shown to be the case for one free-form method using synthetic clusters \citep{gho20}. We note, however, that there remains the possibility that all mass reconstructions may be biased in the same direction.

As we enter this new era of gravitational lensing research, it is important to evaluate the progress of our modelling paradigm. This involves developing effective ways to compare and quantify the similarity between different lens models. This has been done in the past \citep{meneghetti17,acebron17,priewe17,remolina18,raney20} with the Hubble Frontier Fields \citep[HFF, ][]{lotz17}. It has been found that circularly averaged mass profiles agree quite well across different lens modelling methodologies \citep{meneghetti17,raney20}. Magnification maps, however, remain poorly constrained, especially in regions with high magnification, $\mu \gtrsim 10$ \citep{priewe17,raney20}. This is not entirely surprising because the effectiveness of observed images in constraining the lens model has been shown to drive primarily the large scale smooth mass profile \citep{lasko23}, but not necessarily the intermediate and smaller scale structure within the cluster.

Currently, observed multiple image catalogs of cluster-scale lenses have significantly increased thanks to JWST. A prominent example of this is the lens MACS J0416.1-2403 (MACSJ0416) at $z=0.396$. MACSJ0416 currently has the most observed lensed images at 343 total \citep{diego23a} with 303 spectroscopically confirmed \citep{rihtarsic24}, thus making it an ideal candidate to test whether lens models are converging towards a common solution with the increase in image constraints. This lens is part of the HFF, and, as a result of its large image catalog, has been the subject of renewed interest for lens modelling \citep{bergamini23,diego23a,cha23,perera24b,rihtarsic24}. MACSJ0416 is an actively merging cluster with a typical bimodal and elongated mass structure \citep{zitrin13,jauzac14,jauzac15,balestra16,kawamata16}. As is expected in merging clusters, it is dynamically complex and predicted to have abundant substructures on many length scales \citep{jauzac18,cerini23}, thus making it a subject of interest to constrain the nature of dark matter \citep{natarajan17,caminha17,bonamigo17,bonamigo18,perera24b}. One possible way to do this has recently been suggested by \cite{zimmerman21}, to use observed images of the same source that form close together, indicating proximity to the critical curve.

In this paper, we compare and quantify the similarity between various lens models of MACSJ0416. Given the large number of images that have been discovered in MACSJ0416 over the years, we can assess whether or not lens models have been converging towards a common solution with greater number of images. We study this by using three different comparison metrics, chosen such that our results are not systematically biased. These results serve as an evaluation of the current state of the art lens modelling methodologies by examining whether techniques are achieving the implicit goal of accurately reconstructing gravitational lenses. We emphasize that due to the merging and complex structure of MACSJ0416, the results presented here may not be generally true for lens models of all clusters, especially relaxed ones. In Section \ref{txt:data} we present our sample of lens models. In Section \ref{txt:metrics} we describe the three comparison metrics we use to quantify similarity between lens models. In Section \ref{txt:results} we present results of the similarity between lens models and how they have evolved with number of images. Section \ref{txt:conclusions} presents discussion, interpretation, and suggestions for future work in lens modelling.

\section{Data}\label{txt:data}

\begin{table}
    \centering
	\caption{Lens Reconstructions of MACS J0416.1-2403}
	\begin{tabular}{ccccc} 
		\hline
		Lens Model & Method & $N_{\rm im}$ & $\Delta_{RMS}$ & Reference \\
		\hline
  		Z-NFW-v3 & PIEMD+eNFW (Par) & 94 (A) & 1.37" & 1\\
  		CATS-v4.1 & {\tt LensTool} (Par) & 116 (A) & 0.72" & 2 \\
  		Caminha17 & {\tt LensTool} (Par) & 102 (A) & 0.59" & 3 \\
  		Sharon14 & {\tt LensTool} (Par) & 97 (A) & 0.51" & 4 \\
  		Williams16 & {\tt GRALE} (FF) & 101 (A) & N/A & 5 \\
  		Diego19 & {\tt WSLAP+} (H) & 95 (A) & 0.62" & 6 \\
  		Keeton20 & {\tt Keeton} (Par) & 95 (A) & 0.52" & 7 \\
  		Glafic18 & {\tt glafic} (Par) & 202 (B) & 0.50" & 8 \\
  		Richard21 & {\tt LensTool} (Par) & 198 (B) & 0.58" & 9 \\
  		Bergamini23 & {\tt LensTool} (Par) & 237 (B) & 0.43" & 10 \\
  		Diego23 & {\tt WSLAP+} (H) & 343 (B) & N/A & 11 \\
  		MARS23 & {\tt MARS} (FF) & 236 (B) & 0.08" & 12 \\
  		Perera25 & {\tt GRALE} (FF) & 237 (B) & 0.19" & 13 \\
  		Rihtarsic24 & {\tt LensTool} (Par) & 303 (B) & 0.53" & 14 \\
		\hline
	\end{tabular}\\
\medskip{A summary table of recent lens models for MACS J0416.1-2403. The columns list the following: lens model name that we adopt for this work, the reconstruction method and type (``Par'' for parametric, ``H'' for hybrid, and ``FF'' for free-form), the number of images $N_{\rm im}$ used (with A and B denoting that the model falls in the $< 150$ and $> 150$ image bin, respectively), the lens plane RMS $\Delta_{RMS}$ (if provided), and the lens model's reference. All models use spectroscopically confirmed images except Diego23, which has 237 secure images in the full sample of 343 images. References: (1) \cite{zitrin13}, (2) \cite{jauzac14}, (3) \cite{caminha17}, (4) \cite{johnson14}, (5) \cite{sebesta16}, (6) \cite{vegaferrero19}, (7) \cite{raney20}, (8) \cite{kawamata18} (9) \cite{richard21}, (10) \cite{bergamini23}, (11) \cite{diego23a}, (12) \cite{cha23}, (13), \cite{perera24b}, (14) \cite{rihtarsic24}.}
\label{tab:summary}
\end{table}

MACSJ0416 represents the lensing cluster with the most observed images used for modeling, currently numbering at 343 \citep{diego23a}. The number of identified lensed images has increased substantially over the last decade, with the advent of JWST contributing to an additional $\sim$100 identified images. 

The increasing abundance of images in this lens over time makes it an ideal candidate to test whether its gravitational lens models have been converging to a common solution. Table \ref{tab:summary} lists a variety of published gravitational lens models of MACSJ0416 that we use in this work. In total, we chose a sample of 14 lens models that effectively samples the diversity in modelling methods, number of images used as constraints ($N_{\rm im}$), and lens plane root-mean-square (RMS):
\begin{equation}\label{eq:rms}
    \Delta_{RMS} = \sqrt{\frac{\sum_i^{N_{\rm im}} |\boldsymbol{r_{i,\rm obs}} - \boldsymbol{r_{i,\rm rec}}|^2}{N_{\rm im}}}
\end{equation}
where $\boldsymbol{r_{i,\rm obs}}$ and $\boldsymbol{r_{i,\rm rec}}$ are the observed and reconstructed $i$th image positions, respectively. In our sample, nine models are parametric, three are free-form, and two are hybrid. We define parametric methods to be those that include cluster member properties explicitly as physically motivated priors (both galaxy-scale and cluster-scale properties have parametric forms), while free-form methods do not include any cluster or cluster member information as a prior. Hybrid methods typically combine a free-form basis with parametric cluster member properties. Many reconstruction methods have been used more than once by their teams on the same galaxy cluster: early models with fewer multiple lensed images have been followed up one or a few years later with updated models that use more images as they became available. Most lens inversion methods have also undergone refinements over the years. In our study, we try to choose two models from the same lens inversion method but with different $N_{\rm im}$, such as WSLAP+ with $N_{\rm im}$ of 95 \citep{vegaferrero19} and 343 \citep{diego23a}. It should be noted that the most common reconstruction method used in the literature is {\tt LensTool}, which has been used at many $N_{\rm im}$ steps. We direct the reader to the references listed in Table \ref{tab:summary} for more details on each lens model\footnote{For convenience, all the lens models are gathered and publicly available at \url{https://github.com/derekperera/MACSJ0416-Model-Comparison}}.

In our sample, we have 7 models that utilize $< 150$ images and 7 utilizing $> 150$ images. In comparing between the models in these two bins, we are able to test whether an increase in $N_{\rm im}$ leads the lens models of MACSJ0416 to converge to a common solution. It is important to realize that the source redshift distribution of each model's input image dataset remains remarkably consistent across all models, with a source redshift mode of $3.06 \pm 0.39$. From this, we conclude that differences in the source redshifts of images between models likely have minimal effect on our analysis and results.

Furthermore, it is worth considering the impact of false  images on contributing to differences in mass distribution, such as through the generation of anomalous (light-unaffiliated) substructures. In this context, false images refer to images confirmed to be misidentified in future analyses. False images are rare in MACSJ0416 and when they occur, have produced anomalous substructure on scales $\lesssim 1"$ \citep{cha23}. For this reason, we do not expect false images to contribute to model differences on scales greater than 1".

Instead, it is more common for image catalogs to differ in terms of the confidence level of image identifications. This is a common occurrence in gravitational lens modelling, and in MACSJ0416, $\sim$10\% of images have significant differences in confidence level across models. Likewise we expect low contribution to model differences from images with varying confidence assessments, since the vast majority of images are consistently securely identified. 

The metrics that we use to compare these lens models are described in Section \ref{txt:metrics}. We note that this comparison is restricted to each model's $\kappa$ distribution, which can be obtained from each model's reference in Table \ref{tab:summary}.

\section{Comparison Metrics}\label{txt:metrics}

With our sample of lens models in Table \ref{tab:summary}, we first divide the sample into two bins: (A) models containing $< 150$ images, and (B) $> 150$ images. The cutoff of 150 images is rather arbitrary, however, it represents a natural midpoint in the progression of identified images over time. Importantly, this cutoff allows for equally sized A and B bins, allowing for effective comparison. In general, if models are converging, then the models with $> 150$ images should be more similar to one another than models with $< 150$ images.

With this methodology, we introduce 3 metrics below to quantify the similarity between models. The choice of 3 metrics is used to reduce systematic bias that may exist if we restricted the analysis to a single metric. We emphasize that the 3 metrics we use in this work are not the only ones that can be used to judge similarity of lens models. Our metrics are applied to each model's convergence $\kappa$ maps, since this is directly related to the reconstructed mass profile of the lens. We also apply the three metrics only within a predefined strong lensing region surrounding the rough extent of the observed images, shown as a green rectangle in the top panel of Figure \ref{fig:PDhist} with length of 63" and width of 118". For each $\kappa$ map, we interpolate each model onto a standard grid with 1.41 arcsec per pixel. This standard grid is centered on a zero point at RA, Dec = 64.037, -24.071 degrees. We discuss the sensitivity of our metrics to this resolution in Appendix \ref{txt:sensitive}. We also perform a simple test to see if free-form models (particularly {\tt GRALE}) may be overfitting, in Appendix \ref{txt:overfit}. We emphasize that our comparisons are designed to directly compare the reconstructed mass distributions of different lens models, regardless of the degrees of freedom, predictive power, and other technicalities of individual methods.

\subsection{Median Percent Difference}\label{txt:PD}
The median percent difference (MPD) is an intuitive and commonly used metric to quantify the similarity between lens models \citep{bergamini23}. For each model comparison in each bin (A or B), we first calculate the percent difference map at each pixel location in the standard grid. Thus, the percent difference (PD) at a given location $\boldsymbol{\theta}$ is:
\begin{equation}
    PD(\boldsymbol{\theta}) = 2\frac{|\kappa_1(\boldsymbol{\theta}) - \kappa_2(\boldsymbol{\theta})|}{\kappa_1(\boldsymbol{\theta}) + \kappa_2(\boldsymbol{\theta})} \times 100
\end{equation}\label{eq:PD}
where $\kappa_1$ and $\kappa_2$ are the two models being compared.

\begin{figure}
\includegraphics[trim={1.5cm 0.35cm 1.5cm 0.35cm},clip,width=0.49\textwidth]{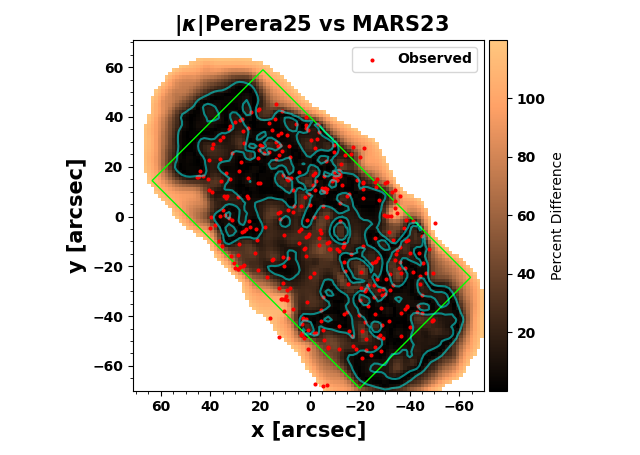}
\includegraphics[trim={0cm 0cm 0cm 0cm},clip,width=0.49\textwidth]{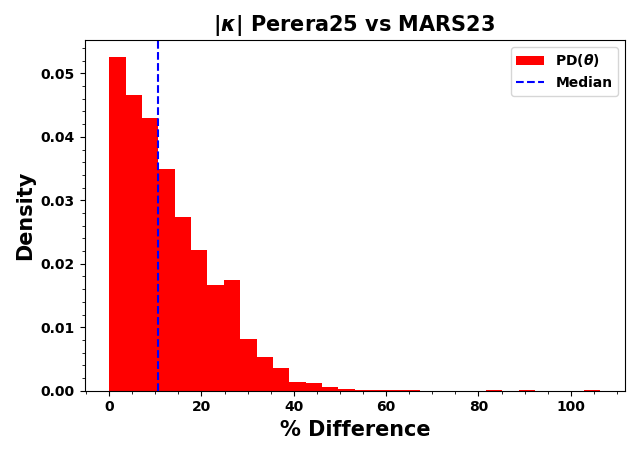}
\caption{{\it Top:} Percent Difference map of MACSJ0416 between the models Perera25 \citep{perera24b} and MARS23 \citep{cha23}. Darker shaded regions correspond to locations where these two models have similar $\kappa$. The lensing region is bound by the bright green rectangle, and roughly surrounds the vast majority of the multiple images (red dots). Regions further away from the lensing region have large PD $> 120\%$. These regions are colored in white to make the lensing region structure more apparent. The turquoise contour indicates a PD of $10\%$. The center of the grid at (0,0) corresponds to RA,Dec = 64.0373,-24.071 degrees. {\it Bottom:} Histogram of the percent difference (PD) calculated at each pixel within the green rectangle from the upper panel. The blue dashed line indicates the median percent difference, equivalent to $10.6\%$ in this case. }
\label{fig:PDhist}
\end{figure}

Calculating this for all $\boldsymbol{\theta}$ in the standard grid gives us an array of PDs for a single comparison, which we can plot as a histogram as in the bottom panel of Figure \ref{fig:PDhist}. To quantify the similarity between $\kappa_1$ and $\kappa_2$, we take the median of this array. As mentioned above, we restrict the calculation of the PDs to a rectangular region surrounding most of the images, shown as a green rectangle in the top panel of Figure \ref{fig:PDhist}. A smaller median percent difference value in this region implies that $\kappa_1$ and $\kappa_2$ are more similar to one another. Figure \ref{fig:PDhist} shows one example comparison for visualization purposes of this metric. We show additional example comparisons for different reconstruction methods in Appendix \ref{txt:morecompa}.

\subsection{Frechet Distance} \label{txt:FD}

\begin{figure}
\includegraphics[trim={4.7cm 0cm 6.0cm 0cm},clip,width=0.49\textwidth]{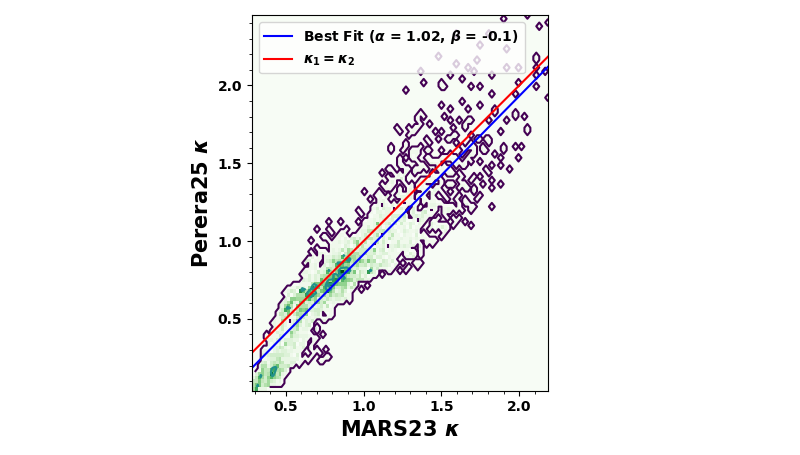}
\caption{Normalized 2D histogram of the convergence values between models Perera25 \citep{perera24b} and MARS23 \citep{cha23}. Green color indicates a greater concentration of convergence points. The purple contour denotes the boundary where there is no corresponding dataset on the histogram.} The red line denotes the ideal scenario at which both models have the same convergence. The blue line is the best fit linear line that fits the histogram. The metric $\delta_F$ described in Section \ref{txt:FD} is defined as the Frechet distance between the red and blue lines. For this comparison, $\delta_F = 0.103$. The portrayal here as a normalized histogram is to more easily show the visual comparison between the two maps; the fit is not calculated on the normalized histogram. Furthermore, additional comparison examples are shown in Appendix \ref{txt:morecompa}.
\label{fig:FD}
\end{figure}

Another metric we incorporate into this analysis makes use of the Frechet distance $\delta_F$. The Frechet distance is a similarity measure between two curves which has been used as a comparison metric in studies of radio galaxies \citep{slijepcevic22} and generating galaxy images \citep{holzschuh22}. For our purposes, it can be simplified into a three step process.

First, we take the convergence values for the two models being compared ($\kappa_1$ and $\kappa_2$) within the lensing region. We then plot $\kappa_1$ versus $\kappa_2$, as shown as a normalized 2D histogram in Figure \ref{fig:FD}. Second, we find the linear line of best fit between $\kappa_1$ and $\kappa_2$:
\begin{equation}\label{eq:kappabest}
    \kappa_2 = \alpha\kappa_1 + \beta
\end{equation}
where $\alpha$ and $\beta$ are the linear best fit parameters. We emphasize that this fit is between $\kappa_1$ and $\kappa_2$ at each point in the standard grid. 

Finally, we calculate the Frechet distance between this best fit line and the curve $\kappa_1 = \kappa_2$. If $\kappa_1$ and $\kappa_2$ are similar to one another, then the best fit line between them should be close to $\kappa_1 = \kappa_2$. Therefore, the Frechet distance serves as a way to quantify how close the best fit line is to $\kappa_1 = \kappa_2$. Since both these curves are linear, the Frechet distance simplifies to just the maximum perpendicular distance between the two curves:
\begin{equation}\label{eq:frechet}
    \delta_F = \max \{|\left(\alpha-1\right)\kappa_1 + \beta|\}
\end{equation}
where we restrict $\kappa_1 \leq 3$ since convergence values typically do not surpass this limit. A smaller $\delta_F$ value indicates that the models being compared have a higher level of similarity. Figure \ref{fig:FD} illustrates the process outlined here. Because the calculation of $\delta_F$ requires a linear fit to the $\kappa_1$ versus $\kappa_2$ relation, this metric is dependent on the resolution of the lens model. We discuss the implications of this sensitivity in Appendix \ref{txt:sensitive}.

We note that it is possible to swap $\kappa_1$ and $\kappa_2$ in equation \ref{eq:kappabest}, which would result in slightly different $\alpha$ and $\beta$, and accordingly a different $\delta_F$. Therefore, to be consistent with this metric, we quote the smaller value of $\delta_F$ between the two scenarios. While this is not a complete fix, in practice it does not significantly change the overall results because we do this consistently for each comparison and the linear fits between models have mean $\alpha = 0.97 \pm 0.09$. Since the distribution of the $\alpha$ fit parameter is very close to 1, we conclude that our choice for consistency of this metric does not significantly change the overall result. The tight distribution of $\alpha$ around 1 means that $\beta$ is the main parameter of interest when calculating $\delta_F$. We note that this can allow $\delta_F$ to function as a test of the mass sheet degeneracy, as we elaborate in Appendix \ref{txt:sensitive}. 

\subsection{Wasserstein Distance}\label{txt:WD}
The Wasserstein distance, or more commonly the Earth Movers Distance, is a solution to the optimal transport problem in mathematics, which quantifies the minimum cost in transforming one probability distribution into another. The cost mathematically corresponds to a function relating this transport between the two probability distributions. In our case, we can think of the Wasserstein distance as the optimal way to transform $\kappa_1$ into $\kappa_2$. Thus, in this context it is a measure of similarity between convergence maps. This metric has been used in cosmology to discriminate between different simulated cosmological models \citep{tsizh23}, representing a similar astrophysical application of this method to quantify similarity between models. In this section we briefly summarize the method with which we use to calculate the sliced Wasserstein distance. We direct the reader to \cite{bonneel15} for greater detail.

In general, the p-Wasserstein distance is written as \citep{bonneel15}:
\begin{equation}
    W_p\left(\kappa_1,\kappa_2\right) = \left( \inf_{\gamma\in \Gamma\left(\kappa_1,\kappa_2\right)} \int_{\mathbb{R}^2\times\mathbb{R}^2} ||x-y||^p d\gamma\left(x,y\right)\right)^{\frac{1}{p}}
\end{equation}
where $x \in \kappa_1$, $y \in \kappa_2$, $\Gamma\left(\kappa_1,\kappa_2\right)$ is the set of all transportation paths between $\kappa_1$ and $\kappa_2$, and $\gamma \in \Gamma\left(\kappa_1,\kappa_2\right)$, where $\gamma$ is a single path. Basically, for a given mapping between $\kappa_1$ and $\kappa_2$, one calculates the integral of the cost function $||x-y||^p$, representing how much $\kappa_1$ and $\kappa_2$ differ. With this resulting set, its infimum is the Wasserstein distance to $p$th order. To penalize any potential outlier $\kappa$ values, we use $p = 2$.

Evaluating $W_p\left(\kappa_1,\kappa_2\right)$ is computationally expensive, so we instead calculate the sliced Wasserstein distance \citep{bonneel15}:
\begin{equation}
    \overline{W}\left(\kappa_1,\kappa_2\right) = \int_{\mathbb{S}^{d-1}} W_p\left(\mathcal{R}\kappa_1,\mathcal{R}\kappa_2\right) d\theta
\end{equation}
where $\mathbb{S}^{d-1}$ is the unit sphere in real space $\mathbb{R}^d$ such that $\int_{\mathbb{S}^{d-1}} d\theta = 1$, $d$ is the dimensionality, and $\mathcal{R}$ is the Radon transform. The radon transform is an integral transform of a function onto a two-dimensional grid of lines, where the line integral of each line is its value. Essentially, the sliced Wasserstein distance performs a linear projection with the Radon transform for both $\kappa_1$ and $\kappa_2$. It is then on this projection that we calculate the Wasserstein distance and integrate it on the unit sphere. The ``sliced'' naming convention originates from the calculation on the grid of lines from the Radon transform.

For our use in this work, we utilize the software {\tt POT} \citep{flamary21} to calculate $\overline{W}\left(\kappa_1,\kappa_2\right)$. In this code, the calculation of the integral on the unit sphere utilizes a Markov Chain Monte Carlo (MCMC). Therefore, for our results we quote the mean of 1000 projections\footnote{For this work, the distribution of projections is consistently Gaussian, so there is no discernible difference in choosing to quote the mean or median. For simplicity we choose the mean.} of $\overline{W}\left(\kappa_1,\kappa_2\right)$. The smaller the value of $\overline{W}\left(\kappa_1,\kappa_2\right)$, the more similar $\kappa_1$ and $\kappa_2$ are.

\section{Results}\label{txt:results}

\subsection{Full Sample}

\begin{figure*}
\includegraphics[trim={0cm 0cm 0cm 0cm},clip,width=0.32\textwidth]{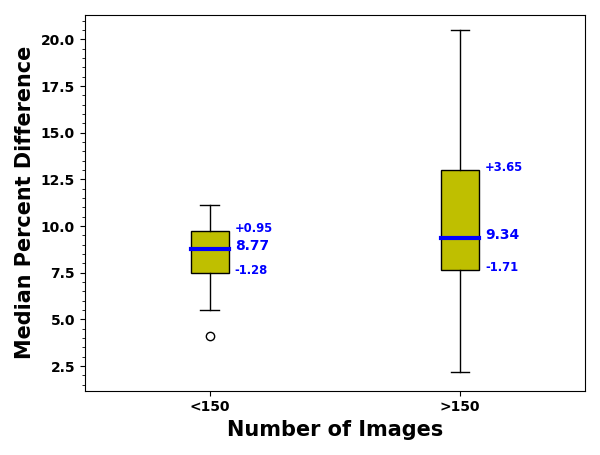}
\includegraphics[trim={0cm 0cm 0cm 0cm},clip,width=0.32\textwidth]{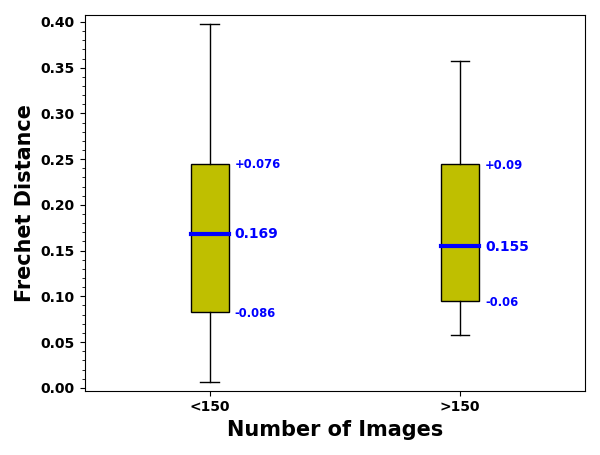}
\includegraphics[trim={0cm 0cm 0cm 0cm},clip,width=0.32\textwidth]{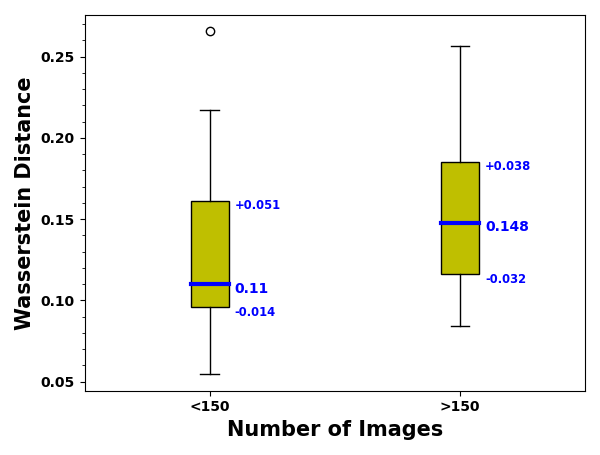}
\caption{ Boxplots for the two image bins of MPD ({\it Left}), $\delta_F$ ({\it Middle}), and $\overline{W}$ ({\it Right}). These comparisons make use of the full sample of lens models described in Table \ref{tab:summary}. The yellow shaded regions depict the interquartile range (middle 50\%) while the blue line depicts the median for each comparison metric.The blue text indicates the values of the median and extent of the interquartile range. The whiskers extend to $1.5\times$ the interquartile range. Circles mark outlier points. The change in distribution of these metrics from $< 150$ to $> 150$ images indicates how $N_{\rm im}$ has influenced similarity among lens models.}
\label{fig:binary}
\end{figure*}

\begin{figure*}
\includegraphics[trim={0cm 0cm 0cm 0cm},clip,width=0.32\textwidth]{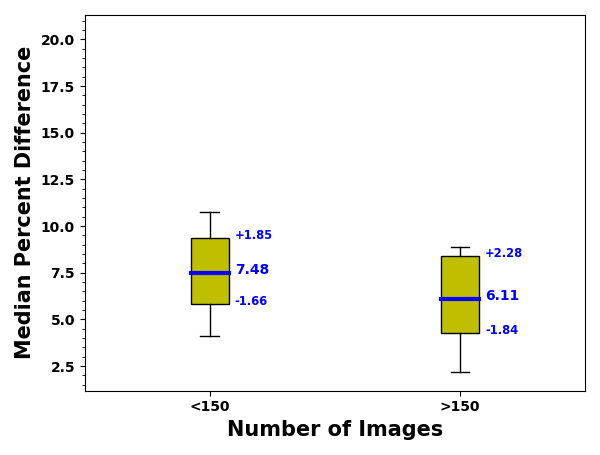}
\includegraphics[trim={0cm 0cm 0cm 0cm},clip,width=0.32\textwidth]{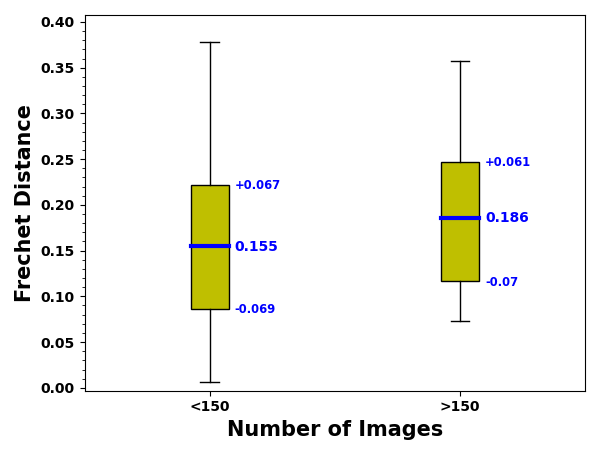}
\includegraphics[trim={0cm 0cm 0cm 0cm},clip,width=0.32\textwidth]{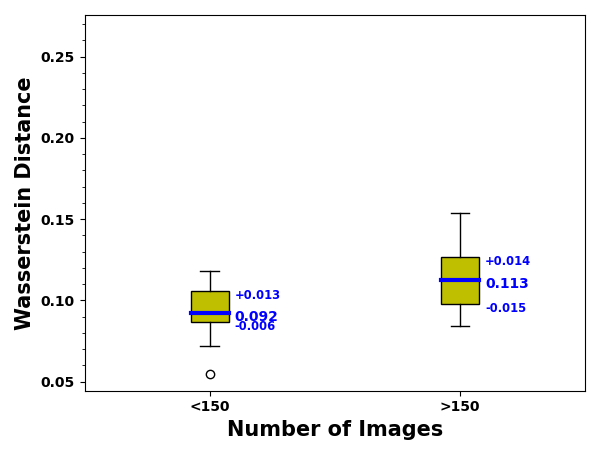}
\caption{Same as Figure \ref{fig:binary} but restricting only to comparisons between the nine parametric models. The vertical range in each metric is fixed to that of the full sample.}
\label{fig:param}
\end{figure*}

\begin{figure*}
\includegraphics[trim={0cm 0cm 0cm 0cm},clip,width=0.32\textwidth]{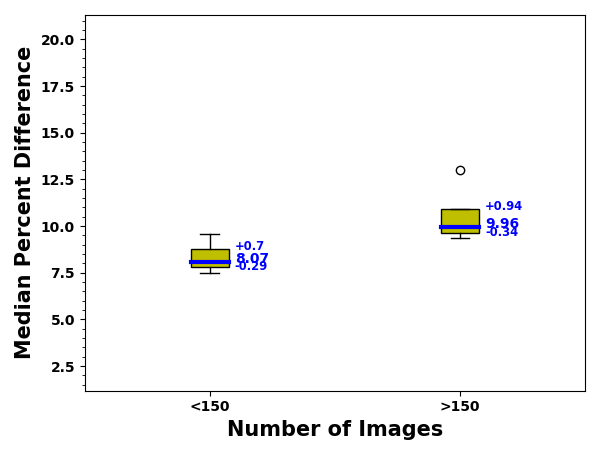}
\includegraphics[trim={0cm 0cm 0cm 0cm},clip,width=0.32\textwidth]{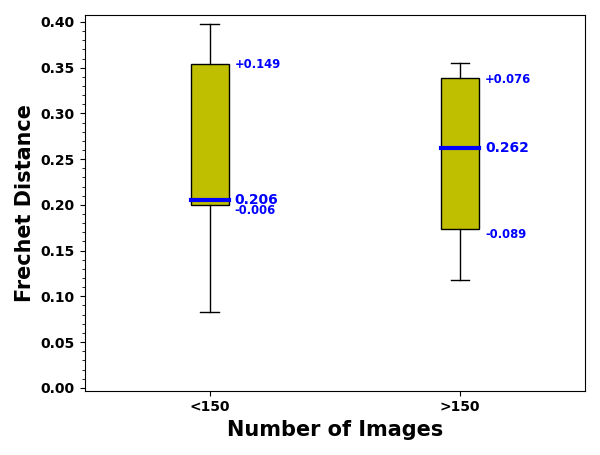}
\includegraphics[trim={0cm 0cm 0cm 0cm},clip,width=0.32\textwidth]{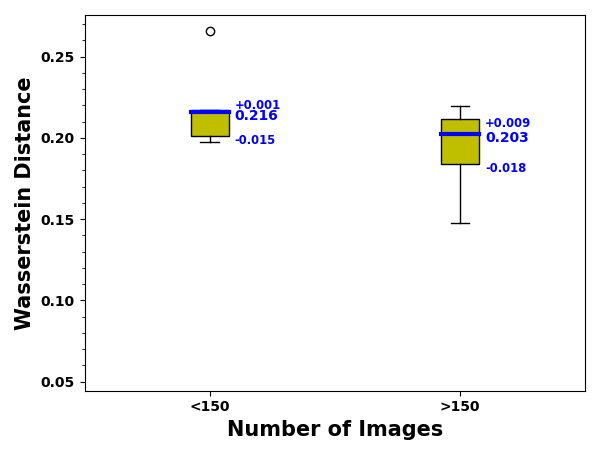}
\caption{Same as Figure \ref{fig:binary} but restricting only to comparisons between parametric models and {\tt GRALE}. The range in each metric is fixed to that of the full sample. We caution that these results use few points, only 5 and 4 for $< 150$ and $> 150$ images, respectively.}
\label{fig:pvff}
\end{figure*}

Following our analysis plan in Section \ref{txt:metrics}, we calculate MPD, $\delta_F$, and $\overline{W}$ for each comparison between each model in their respective image bins, A ($<150$ images) or B ($>150$). Since our sample has 7 lens models in each bin, this amounts to 21 comparisons in both bins for each of the 3 comparison metrics. Quantitatively, we assign the threshold for convergence and divergence of models to be (very roughly) a $\sim1\sigma$ decrease and increase in the median metric values, respectively. We show summary statistics of the full sample for each metric as boxplots in Figure \ref{fig:binary}. For our results, the error in each metric is quoted as the extent of the interquartile range (IQR) for each distribution, corresponding to the middle 50\%.

For the MPD metric, the median does not change significantly with $N_{\rm im}$, only slightly increasing from $8.77^{+0.95}_{-1.28}\%$ to $9.34^{+3.65}_{-1.71}\%$. The main difference between the two bins is that the range increases substantially for $N_{\rm im} > 150$. The lowest MPD is the comparison between Bergamini23 and Rihtarsic24 at $2.20\%$, while the highest is between MARS23 and Glafic18 at $20.51\%$. This result is as expected, since Bergamini23 and Rihtarsic24 both utilize parametric methods (in this case {\tt LensTool}) for their models, with the primary difference being the inclusion of JWST identified images in the latter. Likewise, MARS23 and Glafic18 use free-form and parametric methods, respectively. The two methods are expected to differ due to their differing modelling approaches, with parametric models typically directly modelling cluster member components with explicit mass functions. This result supports the efficacy of MPD in adequately comparing lens models. In fact, the median MPDs are similar to the $\sim10\%$ difference found by \cite{meneghetti17} between different modelling paradigms. The small MPD found between models is consistent with lens models being able to predict source plane redshifts \citep{remolina18}.

For $\delta_F$, the result similar to that of the MPD, with the median $\delta_F$ slightly decreasing from $0.169^{+0.076}_{-0.086}$ to $0.155^{+0.090}_{-0.060}$ in the image bins. The minimum $\delta_F$ is for Sharon14 versus Z-NFW-v3 at 0.007, and the maximum is for Keeton20 versus Williams16 at 0.398. Similar to MPD, $\delta_F$ also reproduces the expected result of dissimilarity between parametric methods and {\tt GRALE}. 

Finally, $\overline{W}$ gives similar results as the other two metrics, with median $\overline{W}$ more clearly increasing with $N_{\rm im}$ from $0.110^{+0.051}_{-0.014}$ to $0.148^{+0.038}_{-0.032}$. The lowest and highest $\overline{W}$ are for Sharon14 versus Caminha17 at 0.054, and Keeton20 versus Williams16 at 0.266. 

In summary of these results for the full sample, we find that all three metrics are remarkably consistent with one another, with no signifcant consistent trend. This gives strong evidence that with our full sample of metrics we are able to draw reasonable conclusions as to whether or not lens models are converging. Towards this question, the MPD and $\overline{W}$ metrics seem to suggest that lens models of MACSJ0416 are slightly diverging in similarity despite the increase in $N_{\rm im}$, while $\delta_F$ suggests the opposite. The most likely conclusion is that there is no trend between bin A ($<150$ images) and bin B ($>150$ images). A statistical test confirms this conclusion: a null hypothesis that there is no trend cannot be rejected as $p$-values are large: 0.21, 0.91, and 0.26 for MPD, $\delta_F$, and $\overline{W}$, respectively. Therefore, we conclude that based on these results, lens models for MACSJ0416 have neither converged nor diverged despite the increase in $N_{\rm im}$.

\subsection{Parametric Only}\label{sec:param}

Having established that lens models of MACSJ0416 are neither converging nor diverging, we can check if this result is influenced by reconstruction method. Figure \ref{fig:param} shows the same image bins, but now restricted to comparisons between parametric models only. Bins A and B have 5 and 4 models each, and so 10 and 6 comparison pairs. We find that in general parametric models are more similar with one another than other methods, evidenced by most of the metrics having lower medians than the full sample, with the only exception being bin B comparisons with $\delta_F$. This is as expected due to the direct modelling of cluster members that is common in parametric methods, as well as the use of standard elliptical parametric forms for cluster-scale mass distribution. There again is not a common trend in each metric with $N_{\rm im}$, with MPD slightly decreasing from $7.48^{+1.85}_{-1.66}\%$ to $6.11^{+2.28}_{-1.84}\%$, $\delta_F$ increasing from $0.155^{+0.067}_{-0.069}$ to $0.186^{+0.061}_{-0.070}$, and $\overline{W}$ increasing from $0.092^{+0.013}_{-0.006}$ to $0.113^{+0.014}_{-0.015}$. This result is also not statistically significant with $p$-values of 0.24, 0.56, and 0.06 for MPD, $\delta_F$, and $\overline{W}$, respectively. Thus, the null hypothesis stating there is no trend cannot be ruled out. For these reasons, we find that parametric models do not bias the results of the full sample and that parametric models have neither converged nor diverged with $N_{\rm im}$. We note that this result is not as robust due to there only being 10 and 6 comparisons in the $<150$ (A) and $>150$ (B) image bins, respectively.

\subsection{Parametric vs. {\tt GRALE}}\label{sec:paramgrale}

In general, we also want to evaluate comparisons between parametric and free-form models. However, since {\tt GRALE} is the only free-form model in our sample with $< 150$ images, we restrict these parametric versus free-form comparisons to those only with {\tt GRALE}. Figure \ref{fig:pvff} shows these results, with each comparison pair having one parametric and one {\tt GRALE} method.

As expected, the metric values are generally higher than those among just parametric models, indicating that the two mass maps in each pair are quite different. The results are very similar to the parametric comparisons of Figures \ref{fig:binary} and \ref{fig:param}, with no common trend among the methods with $N_{\rm im}$. Here,  MPD increases from $8.07^{+0.70}_{-0.26}\%$ to $9.96^{+0.94}_{-0.34}\%$, $\delta_F$ increases from $0.206^{+0.149}_{-0.006}$ to $0.262^{+0.076}_{-0.089}$, and $\overline{W}$ decreases from $0.216^{+0.001}_{-0.015}$ to $0.203^{+0.009}_{-0.018}$. The $p$-values are 0.03, 0.99, and 0.22 for  MPD, $\delta_F$, and $\overline{W}$, respectively. Interestingly, the $p$-value for MPD seems to indicate a trend that the models have diverged. However, we emphasize that these results are also not robust as there are only 5 and 4 comparisons in the < 150 and > 150 image bins, respectively. For this reason, and since $\delta_F$ and $\overline{W}$ remain statistically insignificant, we conclude that parametric vs. free-form comparisons have not biased our results, and do not find that these have either converged or diverged.

Ideally we would like to also evaluate any possible trend between free-form and hybrid models. Unfortunately, there is only 1 and 2 comparisons available in the $<150$ and $>150$ image bins, respectively. This is far too small of a sample to derive summary statistics. Therefore, with our current sample we are unable to establish if any trend exists between free-form and hybrid models. This would be a natural next step to evaluate in the future with more free-form and hybrid models that will be generated.

\begin{figure*}
\includegraphics[trim={0cm 0cm 0cm 0cm},clip,width=0.32\textwidth]{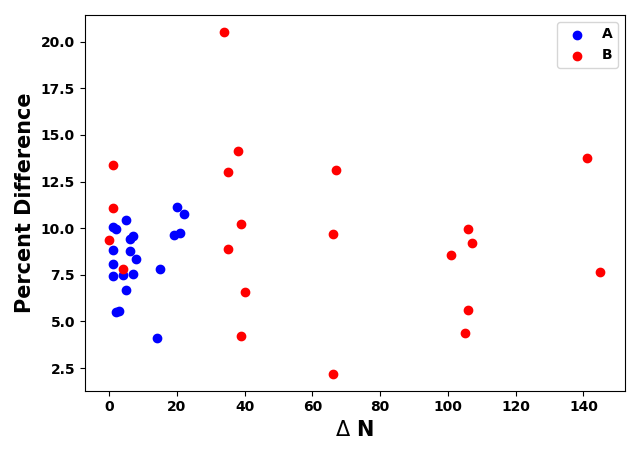}
\includegraphics[trim={0cm 0cm 0cm 0cm},clip,width=0.32\textwidth]{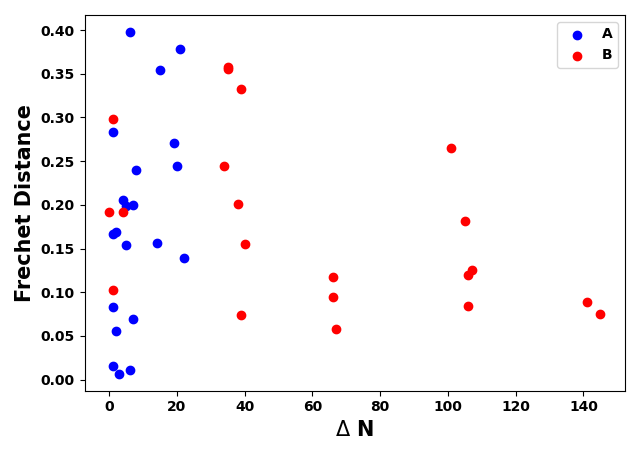}
\includegraphics[trim={0cm 0cm 0cm 0cm},clip,width=0.32\textwidth]{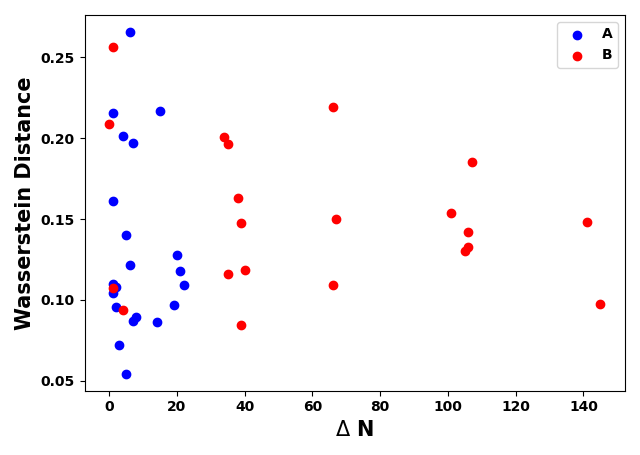}
\caption{Individual comparisons between models within the two image bins of MPD ({\it Left}), $\delta_F$ ({\it Middle}), and $\overline{W}$ ({\it Right}) as a function of the image difference between the compared models ($\Delta N$). Blue points are for comparisons between models in bin A, while red points are for comparisons between models in bin B.}
\label{fig:altfigs1}
\end{figure*}

\begin{figure*}
\includegraphics[trim={0cm 0cm 0cm 0cm},clip,width=0.32\textwidth]{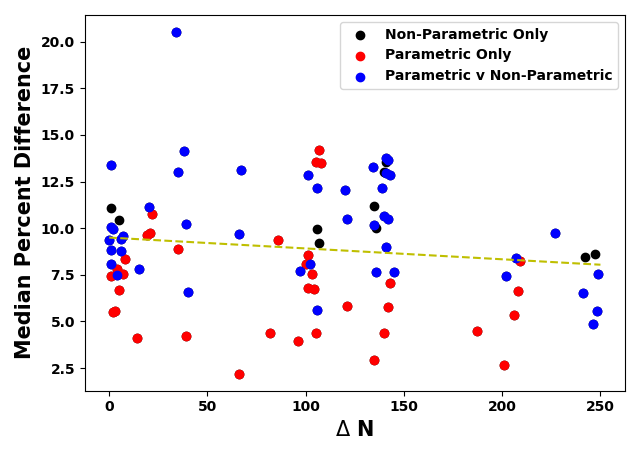}
\includegraphics[trim={0cm 0cm 0cm 0cm},clip,width=0.32\textwidth]{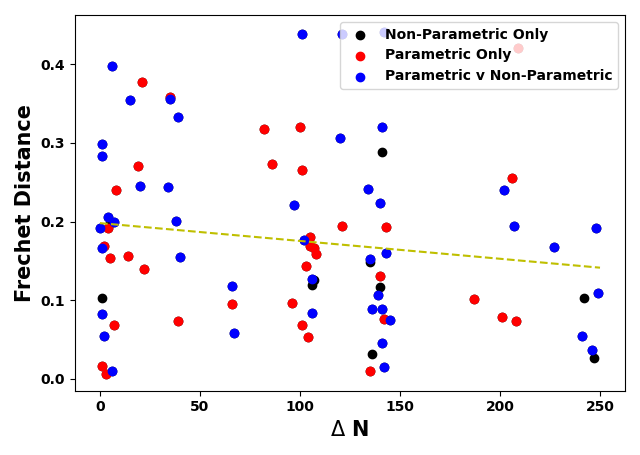}
\includegraphics[trim={0cm 0cm 0cm 0cm},clip,width=0.32\textwidth]{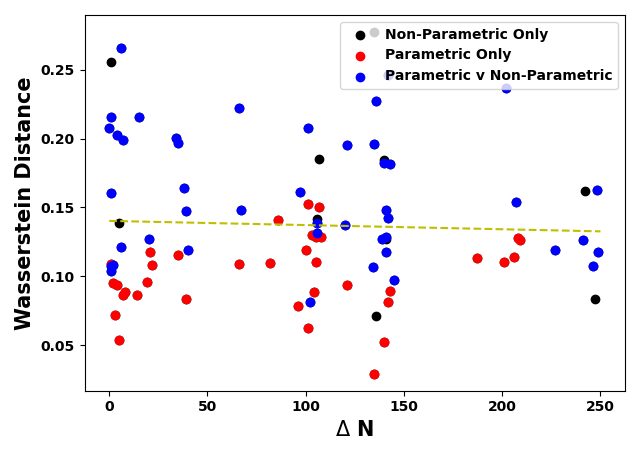}
\caption{Individual comparisons between all models, independent of the two image bins, for MPD ({\it Left}), $\delta_F$ ({\it Middle}), and $\overline{W}$ ({\it Right}) as a function of the image difference between the compared models ($\Delta N$). Red, blue, and black dots indicate comparisons between only parametric models, parametric and non-parametric models, and only non-parametric models, respectively. The dashed yellow line indicates the linear best fit line.}
\label{fig:metrictrend}
\end{figure*}

\subsection{Alternative Representations of Lens Model Trends}

It is important to consider whether our current lens model sample division into the two image bins is sufficiently capturing the existing internal trends between lens models. This consideration is motivated by the fact that the range of $N_{\rm im}$ in bin B is 145, which is significantly larger than the range of 22 in bin A. This discrepancy necessitates that we check whether our division into two image bins is sufficient to conclude that models have neither converged nor diverged.

To test this, we show an alternative representation of the comparisons between models in Figure \ref{fig:altfigs1}, where comparisons between models using each metric are shown as a function of the image difference between compared models ($\Delta N$) in each bin. This figure can be thought of as an alternative representation of Figure \ref{fig:binary}. As can be seen, bin B contains comparisons with  $\Delta N > 100$, capturing the wide range of $N_{\rm im}$ existing within it. Any trend within bin B would manifest as the respective metric varying significantly with $\Delta N$. However, as we can see, there is considerable scatter with each metric as $\Delta N$ increases, and the overall metric trend remains flat. Therefore, we conclude that there is no internal trend within either image bin, consistent with our original conclusion.

For completeness, we also show the same plots as Figure \ref{fig:altfigs1} but ignoring the image bins. In Figure \ref{fig:metrictrend}, comparisons between models in the two image bins are also included, which is why $\Delta N$ extends to $>200$. We note that this is an imperfect representation of the original goal of this paper, as $\Delta N$ is not a measure of time (i.e. a given value of $\Delta N$ can be comparisons between models from either bin A or B). Nevertheless, this representation validates the findings from Figure \ref{fig:altfigs1}, namely that there does not appear to be any significant trend for any metric and $\Delta N$. This is quantified with a best fit line in each case, where we find slopes of $-0.0058 \pm 0.0045$, $-0.0002 \pm 0.0001$, and $-3.0326 \pm 7.2656 \times 10^{-5}$, for MPD, $\delta_F$, and $\overline{W}$, respectively. All three metric trends find a slope $\ll 1$, indicating a very flat trend with $\Delta N$. Furthermore, the Pearson correlation coefficient is -0.134, -0.152, and -0.044 for MPD, $\delta_F$, and $\overline{W}$, respectively, indicating weak correlation. All of this supports the conclusion that there are no internal trends between the models. Therefore, splitting the sample into bins A and B seems to be the best way to examine the convergence of models.

\section{Discussion and Conclusions}\label{txt:conclusions}

We present three comparison metrics capable of evaluating whether or not lens models of MACSJ0416 are converging or diverging with the increased number of images, $N_{\rm im}$. Generally, one expects that an increase in observed images should lead to a more accurate mass reconstruction, and hence convergence of the mass maps based on different lens inversion methods. However, using a sample of 14 lens models, we find that lens models of MACSJ0416 are neither converging nor diverging despite the drastic recent increase in $N_{\rm im}$.  Importantly, this result does not seem to be biased by the reconstruction method.

There are interesting interpretations and consequences of this result. Foremost is that even though recent lens models utilizing more images have been finding lower $\Delta_{RMS}$ \citep[e.g.][]{bergamini23,cha23,perera24b}, their mass distributions have not trended to a common solution. This suggests that lensing degeneracies are not being sufficiently broken by increased $N_{\rm im}$. This may not be a very surprising result given that lensed images alone cannot always adequately constrain deviations from cluster-scale smooth elliptical density profile on scales spanning a few to tens of arcseconds \citep{lasko23}, thus requiring alternative probes. Similarly, in parametric modelling techniques, masses of cluster member galaxies are known to be degenerate with the smooth cluster component, unless spectroscopy is used to measure velocity dispersions \citep{limousin16,bergamini19,limousin22,beauchesne24}. Therefore, new lens models of clusters need to take greater care in accounting for degeneracies. 

The primary degeneracy of concern is the shape degeneracy, defined as the fact that different realizations of mass distributions can fit the data equally well. Since circularly averaged mass profiles of different models agree quite well on scales $>100$ kpc \citep{meneghetti17}, the shape degeneracy manifests as variations in the physical shape of the mass distribution (and consequently the critical curve and magnification map) on sub-cluster scales. This is easily seen as the dominant source of dissimilarity between models in Figure \ref{fig:PDhist}, where regions of greater percent difference are randomly dispersed throughout the cluster.

Attempting to break shape degeneracies in these models is likely to require observational constraints in addition to increased $N_{\rm im}$. For example, in MACS J1149, systematic biases in the lens models are ruled out using observed flux ratios \citep{hwilliams24}. In this case the lensed sources are HII regions, which are standardizable candles through the empirical $L-\sigma$ relation \citep{melnick88}. \cite{hwilliams24} is able to show the value of using flux ratios as an effective way to assess the accuracy of lens models. Therefore, we propose explicitly including flux ratios in the future as model constraints to reduce systematic bias and hopefully work towards breaking shape degeneracies. In fact, recent work from \cite{mediavilla24} showed that incorporating flux ratios into lens models can help break shape degeneracies in some cases by constraining the radial mass distribution. Building off of tests such as this to more complex cluster scale lens models is a necessary next step in evaluating the value of flux ratios in the lens modelling process. It is important to note that incorporating flux ratios can only address shape degeneracies, and does not themselves help to break the mass sheet degeneracy.

In general, the use of flux ratios as a constraint in lens models is underutilized due to existing challenges. For compact sources such as stars, binaries, quasars, or Type Ia supernovae, it is typically difficult to determine whether flux ratios of lensed images are a result of differences in magnification from the lens model, variability in the source, or microlensing, especially if the images are close to a critical curve. The easiest way to determine if intrinsic source variability or microlensing are an issue is to acquire spectroscopic observations of lensed sources, allowing accurate identification of the source and thus its intrinsic luminosity. These sources are commonly observed in cluster-scale lenses, such as with JWST's Prime Extragalactic Areas for Reionization and Lensing Science \citep[PEARLS, ][]{windhorst23} program. The knowledge of the intrinsic luminosity of a source allows the use of flux ratios as a direct constraint on the lens model to break the shape degeneracy. Since magnification maps are known to disagree at $\sim30\%$ for $\mu \lesssim 10$ \citep{rodney15,meneghetti17,priewe17,raney20}, this would be an important constraint in reducing the dispersion in magnification. We note that care must be taken in correcting for any potential remaining source variability and microlensing effects even with a source identification.

In addition, a unique feature of MACSJ0416 is that it is host to a plethora of highly magnified transient events \citep{rodney18,chen19,kaurov19,kelly22,yan23,diego23b}, which have been used to study microlensing statistics \citep{li24}, properties of high redshift stars \citep{diego23b}, and dark matter \citep{diego23b,abe23,perera24b}. 

Because of high magnifications near cluster critical curves, these transients are found within less than an arcsecond from the critical curves, suggesting that one can use transients to accurately outline cluster critical curves, aiding in mass reconstruction on scales much larger than an arcsecond \citep{dai18,zimmerman21} and helping to break the shape degeneracy. 

However, this technique does not always work well. In the Spock arc region in MACSJ0416 the mass structure is quite complicated, and even though there is a large number of transients, there is still ambiguity in the shape of the critical curve \citep{diego24,perera24b}. Another example is the Dragon arc in Abell 370, where different models predict a range of different critical curves \citep{li24,Broadhurst2024,limousin24}. Despite these challenges, the prevalence and increasing sample of transients discovered in lensed arcs near the critical curves has recently opened a new frontier in cluster lensing.

Transients can also be useful in determining the density structure on very small, subarcsecond angular scales. On these scales, standard $\Lambda$CDM predicts the existence of numerous compact dark matter substructures \citep{klypin99,moore99}. Their abundance, and density structure can be constrained using these transient events \citep{dai18}. We note that these subhalos are unlikely to impact the mass distribution on scales spanning a few to tens of arcseconds, since the typical subhalo mass is quite small, $\sim 10^{6-9} M_{\odot}$.


These are just two possible ways to break shape degeneracies in clusters. Both importantly involve utilizing additional constraints as priors in the lens model. It is also important to rigorously evaluate anomalous lens model predictions and practice criticism on model outputs. A common anomalous prediction of lens models are dark matter clumps not associated with any cluster member galaxy or obvious source of mass, which we refer to as light-unaffiliated substructures. Structures such as these are seemingly required by the lens model, although their interpretation is difficult and can be misleading. These have been recovered in both parametric and free-form models of many systems in addition to MACSJ0416 (see below), such as Abell 370 \citep[see Table 2 in][]{ghosh21}, Abell 2744 \citep{jauzac16,furtak23}, Abell 1689 \citep{ghosh23}, RXJ0437 \citep{lagattuta23}, and SDSSJ1004 \citep{perera24a,liesenborgs24}. The existence of light-unaffiliated substructures can be tested in various ways, such as by tracing their correlation with images formed at the maxima of the Fermat potential \citep{ghosh23}, or by making use of property-preserving degeneracies such as the monopole degeneracy \citep{liesenborgs24}. 

Lens models of MACSJ0416 have identified light-unaffiliated substructures across many modelling algorithms \citep{jauzac14,bergamini19,gonzalez20,rihtarsic24,perera24b}, although these detections are not as consistent or ubiquitous as those in Abell 370. While uncommon, these light-unaffiliated substructures have been tested for their existence. For example, \cite{jauzac14} generate an additional model including a potential light-unaffiliated substructure as a prior. The model featured an increased $\Delta_{RMS}$ with the light-unaffiliated substructure, thus suggesting that the feature was not real, at least within the framework of the Lenstool modelling method. Tests such as these offer ways to help determine whether the existence of a light-unaffiliated substructure is supported and can be used as a prior in future models. We note that considering light-unaffiliated substructure does not help to break shape degeneracies, but rather serves as an additional way to constrain the lens model.

It has been argued that parametric models featuring light-unaffiliated dark matter clumps represent a failure to adequately model the lens. In fact, \cite{limousin22} finds that small B-spline perturbations in MACS J1206 are capable of compensating for these features to yield more physically motivated models. This is not a universally true prescription, however, as \cite{limousin24} finds their model of Abell 370 to predict light-unaffiliated substructures that seem to be also commonly reproduced by multiple modelling methods. Due to this ongoing uncertainty, we suggest that all lens models of clusters rigorously identify and evaluate the existence of anomalous clumps.

Returning to MACSJ0416, these suggestions are certainly feasible. Extended sources exist in MACSJ0416 (such as with the Warhol arc), and obtaining spectroscopic data of these sources can help directly constrain flux ratios in a manner similar to the analysis by \cite{hwilliams24}. Similarly, individual subhalos have been constrained with high magnification events \citep{diego23b,abe23,perera24b}, and future analyses can extend this to constrain the subhalo abundance. New JWST observations identified multiply imaged sources \citep{rihtarsic24} place further constraints on allowed positions of dark clumps, which can help evaluate anomalous results from lens models. The hope with all of this is that with the addition of more constraints to lens models not from identified images, shape degeneracies will be able to be broken. Ideally, the breaking of shape degeneracies would reveal that lens models are indeed converging to a common solution, realizing the hope of recovering near-true mass distributions in clusters. Likewise, light-unaffiliated mass substructures recovered in lens models should be tested for their reality, as these can place needed additional constraints on future models.

There is another way to look at our main result, that the models are not converging. A survey of literature (and Table \ref{tab:summary}) show that $\Delta_{RMS}$ has been improving for most models over the years, despite the fact that it is harder to find better solutions with more image constraints. Perhaps the improvement is not completely surprising since the lens models main goal is to minimize this quantity. The uncontested popularity of $\Delta_{RMS}$ is due to the lack of alternative figures of merit, although some others have recently been suggested \citep{chow24}. However, minimizing $\Delta_{RMS}$ for images is not the same as minimizing $\kappa_{\rm model}(\theta)-\kappa_{\rm true}(\theta)$ over the lens plane. Therefore the former does not necessarily lead to the latter. This has been shown to be the case for synthetic clusters \citep{gho20}.  A possible future direction is for lens models to modify the figure of merit to include, in addition to $\Delta_{RMS}$, an explicit requirement that different models converge, using a multi-objective-type approach.

While the results of this paper, that the mass models of cluster lenses from various groups show no sign of converging to a common solution, may appear unexpected, we emphasize that they may not be universally applicable since MACSJ0416 is a merging cluster with a likely complex mass distribution. These results do, however, highlight the need to continue to study lens modelling procedures and degeneracies, especially the shape degeneracy. While it may be possible that we have not reached a critical $N_{\rm im}$ threshold beyond which lens models will begin to converge, we propose that it is necessary to more carefully evaluate our lens models' results, and to continue to compare them to each other. We also suggest further study and observations into alternative underutilized lens model priors in addition to image positions.

\section*{Acknowledgements}
The authors acknowledge the computational resources provided by the Minnesota Supercomputing Institute (MSI), which were essential for this study. DP acknowledges the School of Physics and Astronomy, University of Minnesota for partially supporting this work through the Robert O. Pepin Fellowship. SKL acknowledges the support of RGC/GRF 17312122 issued by the Research Grants Council of Hong Kong SAR. ML acknowledges the Centre National de la Recherche Scientifique (CNRS) and the Centre National des Etudes Spatiale (CNES) for support. The authors would like to thank Alfred Amruth, Jose Diego, Patrick Kelly, and Ashish Kumar Meena for useful discussions and suggestions regarding this work.

\section*{Data Availability}
All the data used in this paper is publicly available and can be conveniently accessed at \url{https://github.com/derekperera/MACSJ0416-Model-Comparison}.

\bibliographystyle{mnras}
\bibliography{references}


\appendix

\section{Appendix A: Sensitivity to Map Resolution of Comparison Metrics}\label{txt:sensitive}

To evaluate the effectiveness of the three comparison metrics that we use (see Section \ref{txt:metrics}), it is important to test the limits of each metric. Since these metrics are numerically evaluated for each model on a standard grid with a resolution of 1.41 arcsec per pixel, this amounts to examining the metric's sensitivity to changes in resolution. 

In general, the use of a higher resolution grid for the lens models helps to resolve finer and smaller scale details in the mass distribution. However, since these small scale mass features are highly subject to shape degeneracies, they have the potential to bias a given comparison metric. Therefore, an effective comparison metric is one that is insensitive to changes in resolution.

To test this, we repeat the calculation of the three metrics we used (as described in Section \ref{txt:metrics}) on a grid with higher resolution of 0.282 arcsec per pixel (originally 1.41 arcsec per pixel). We then quantify the sensitivity to resolution calculating the Pearson correlation coefficient (PCC) between each metric's values on the standard and higher resolution grid. The closer the PCC is to 1, the more insensitive the metric is to resolution, as a perfect linear relation between the metric on either grid implies that resolution does not change the results.

Following this plan, we calculate the PCC between each grid to be 0.999, 0.956, and 0.403 for MPD, $\overline{W}$, and $\delta_F$, respectively. From these results, we can conclude that our MPD and $\overline{W}$ metrics are insensitive to resolution, as they both exhibit very strong correlations. $\delta_F$ still exhibits a high degree of correlation, however it is notably weaker than the other two metrics. The reason for this is systematic, as the calculation of $\delta_F$ requires a linear best fit between the two $\kappa$ distributions being compared (equation \ref{eq:kappabest}). These best fit parameters are inherently dependent on the resolution of the $\kappa$ distributions being compared. Therefore, we conclude that $\delta_F$ is moderately sensitive to resolution.

Even though $\delta_F$ exhibits this sensitivity, we note that it may serve additional purposes apart from quantifying similarity. As we mention in Section \ref{txt:FD} and can be seen in Figure \ref{fig:FD}, the $\alpha$ linear fit parameter is centered tightly around 1. This means that the linear offset parameter $\beta$ dominates in the calculation of $\delta_F$. We can therefore interpret $\delta_F$ as a test of the mass sheet degeneracy between different models. This is because comparisons between models that recover $\alpha \sim 1$ differ primarily due to a constant $\kappa$ offset, akin to a mass sheet. Mass sheet transformation also involves rescaling of the density profile, which affects its slope, but that would be harder to demonstrate, especially with sources that span a range of redshifts. Figure \ref{fig:FD} illustrates this effect, seemingly influenced by the nonlinear skew at low $\kappa$, implying that the differences between Perera25 and MARS23 are the result of the mass sheet degeneracy. We note that the connection of $\delta_F$ with the mass sheet degeneracy is apparent in the standard transformation:
\begin{equation}
    \kappa_{\lambda}\left(\boldsymbol{\theta}\right) = \lambda\kappa_{0}\left(\boldsymbol{\theta}\right) + \left(1-\lambda\right)
\end{equation}
where $\lambda$ is the transformation factor, $\kappa_0$ is the original convergence, and $\kappa_{\lambda}$ is the transformed convergence. The similarity can be seen when comparing with equation \ref{eq:frechet}, where $\alpha$ and $\beta$ correspond to the density rescaling and mass sheet, respectively. In fact, the vast majority of sources used in models from the B bin have $D_{ds}/D_s$ within 20$\%$ of each other, which is apparently not large enough to break the mass sheet degeneracy \citep{liesenborgs12}. We emphasize that further study into this is needed to confirm the validity of $\delta_F$ as a tracer of mass sheet degeneracy.

In summary, our MPD and $\overline{W}$ metrics are the most effective at quantifying similarity due to being almost completely insensitive to resolution. $\delta_F$, on the other hand, exhibits moderate and systematic sensitivity to resolution, however, it may serve as an adequate test of the mass sheet degeneracy. In interpreting our results, we conclude that MPD and $\overline{W}$ are the strongest metrics with which to base our conclusions on. Despite the differences between each metric, we note that all 3 metrics gave similar results on the convergence of lens models, showing that each metric appears to be adequately describing the same phenomenon.

\section{Appendix B: Are Free-form Models Overfitting?}\label{txt:overfit}

\begin{figure}
\includegraphics[trim={1.5cm 0.35cm 1.5cm 0.35cm},clip,width=0.49\textwidth]{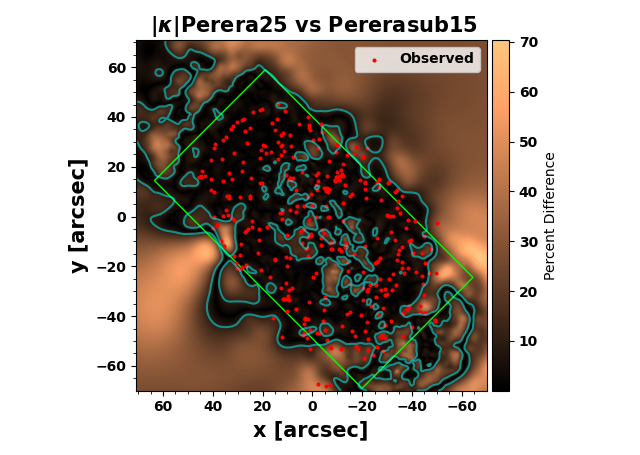}
\includegraphics[trim={0cm 0cm 0cm 0cm},clip,width=0.49\textwidth]{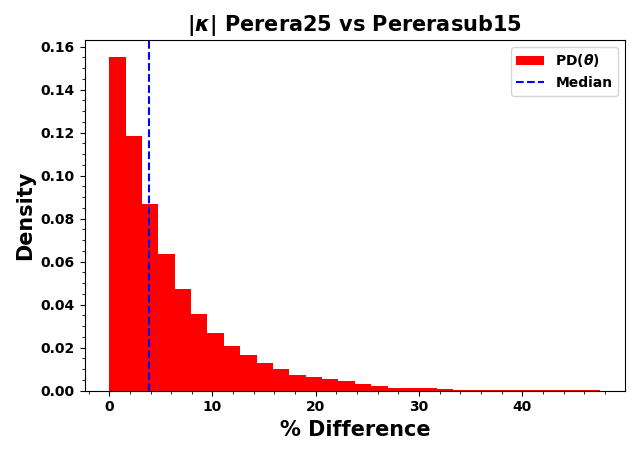}
\caption{Same as Figure \ref{fig:PDhist} but for comparison between Perera25 and Pererasub15. The latter uses 15\% fewer sources as input than Perera25 while continuing to use {\tt GRALE}. The MPD indicated by the blue dashed line is 3.5\% in this case. }
\label{fig:PDoverfit}
\end{figure}

\begin{figure}
\centering
\includegraphics[trim={4cm 0cm 4cm 0cm},clip,width=0.49\textwidth]{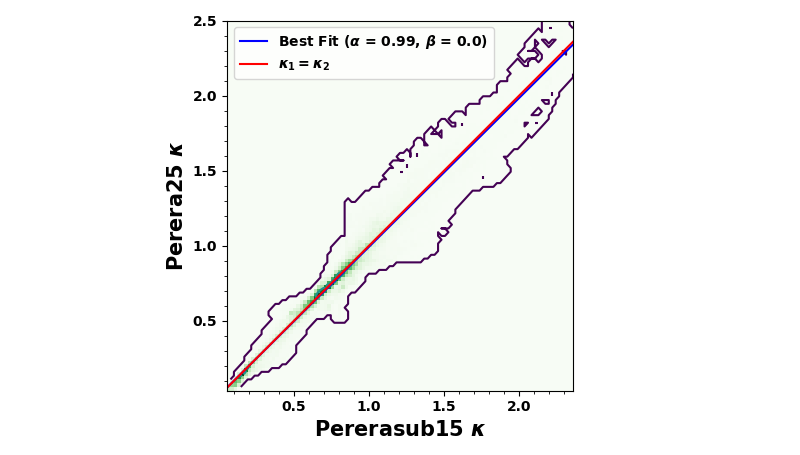}
\caption{Same as Figure \ref{fig:FD} but for Perera25 and Pererasub15. For this comparison, $\delta_F = 0.024$.}
\label{fig:FDoverfit}
\end{figure}

An issue often mentioned in free-form lens modelling is overfitting. This is especially true for models built with noisy data (e.g., underconstrained image positions lacking spectroscopic redshift measurements). The consequence of overfitting would be that these lens models contain erroneous density features that is inflating their dissimilarity with other lens models. This would imply that the observed trend with lens models not converging nor diverging, as discussed in Section \ref{txt:results}, is weak as overfit models contribute to the overall distribution of each metric, with the exception of the parametric-only comparison in Section \ref{sec:param}. Therefore, it is important to test whether overfitting is a potential problem in the free-form models in our lens modelling sample. For this paper, we perform an experiment to check simply if overfitting appears to be a problem in free-form models of MACSJ0416. We note that a full and complete study of overfit lens models is well beyond the scope of this work, and encourage further study on this issue. 

To test overfitting, we use {\tt GRALE} \citep{liesenborgs06,liesenborgs07,liesenborgs20}, a free-form lens inversion methodology utilizing a genetic algorithm to optimize a mass basis on an adaptive grid. The most recent {\tt GRALE} model of MACSJ0416 is presented in \cite{perera24b}, which we have designated as Perera25 in Table \ref{tab:summary}. If {\tt GRALE} were overfitting in Perera25, then the expectation would be that it would fit images better if there were fewer used as input. It would also be expected that a model fit with fewer images would be substantially different from Perera25, or at least have a comparable median metric value with our comparison metrics in Figure \ref{fig:binary}. With this established, we generate a new lens model using {\tt GRALE} that uses 15\% fewer sources than Perera25 as input. This amounts to 215 multiple images used as input, as opposed to 237 as used in Perera25. The excluded sources are randomly selected across all redshifts to reduce bias. The lens model is made in the same way as Perera25, and we direct the reader to \cite{perera24b} for more details on this process. We name this model with 15\% fewer sources: Pererasub15. If Pererasub15 fits images better than Perera25 and is substantially dissimilar according to our metrics, then we can reasonably assume that {\tt GRALE} is overfitting.

Figure \ref{fig:PDoverfit} shows the percent difference map and histogram between Perera25 and Pererasub15. In comparison to the example shown in Figure \ref{fig:PDhist}, we can directly tell that Perera25 and Pererasub15 are much more similar to one another, with the percent difference no greater than $\sim$70\% at any point in the map. The MPD for this comparison in 3.86\%. In Figure \ref{fig:FDoverfit} we show the Frechet distance plot where once again in comparison to the example shown in Figure \ref{fig:FD}, we can see a much stronger similarity between Perera25 and Pererasub15. In this case, $\delta_F = 0.024$. Furthermore, $\overline{W} = 0.034$ in this case, which is lower than any comparison in the full sample. In fact, all three metrics find that Perera25 and Pererasub15 are very similar with one another, and more similar than a typical comparison from our main sample by roughly one standard deviation. Lastly, the $\Delta_{RMS}$ value for Pererasub15 is 0.20", which is slightly larger but still very close to 0.191" which was found for Perera25. Since Pererasub15 is not fitting images better than Perera25 (quantified by $\Delta_{RMS}$) and remains remarkably similar to Perera25, we conclude that {\tt GRALE} is not overfitting in this case.

The exclusion of 22 images in Pererasub15 is sufficient to test overfitting because it is necessary to preserve the input dataset as much as possible. If we instead exclude significantly more sources, it can be argued that comparisons will become meaningless since it would become impossible to distinguish differences between models as stemming from overfitting or substantial input differences. If we temporarily ignore this concern, our conclusions are still supported. Williams16 is the older {\tt GRALE} model of MACSJ0416 that uses 101 images as input, which is a little under half as large as Perera25's input. If we compare these two models we find the MPD, $\delta_F$, and $\overline{W}$ to be $9.01\%$, 0.031, and 0.071, respectively.  Both $\delta_F$ and $\overline{W}$ remain considerably lower than a typical comparison in our main sample, and MPD is very similar to the full sample median MPD. This shows that {\tt GRALE} has not changed any more than any other model comparison. Had Williams16 been substantially dissimilar from Perera25 then we could make the claim that {\tt GRALE} was overfitting as it could be argued that Perera25 would be fitting noise. Since this is not the case we can conclude that {\tt GRALE} is not overfitting.

Lastly, we note that the modelling process for {\tt GRALE} guards against overfitting, perhaps rendering our test unsurprising. {\tt GRALE} operates by subdividing the adaptive grid over many generations, akin to increasing the resolution steadily to improve the fitness. The subdivision step is typically cutoff to prevent the resolution of the adaptive grid to fall below $\sim$1". Even if overfitting were occurring on these scales, all {\tt GRALE} models average over an ensemble of 40 runs, which would eliminate these small scale overfitting features.

If we assume that this same result is true for other free-form lens modelling strategies, then we can rule out overfitting as a major factor in biasing our results. This assumption, importantly, needs to be verified in other free-form modelling algorithms. 


\section{Appendix C: Additional Comparison Plots}\label{txt:morecompa}

\begin{figure}
\includegraphics[trim={1.5cm 0.35cm 1.5cm 0.35cm},clip,width=0.49\textwidth]{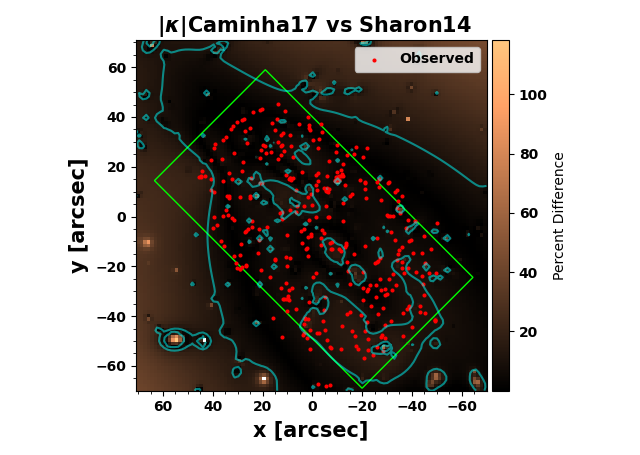}
\includegraphics[trim={0cm 0cm 0cm 0cm},clip,width=0.49\textwidth]{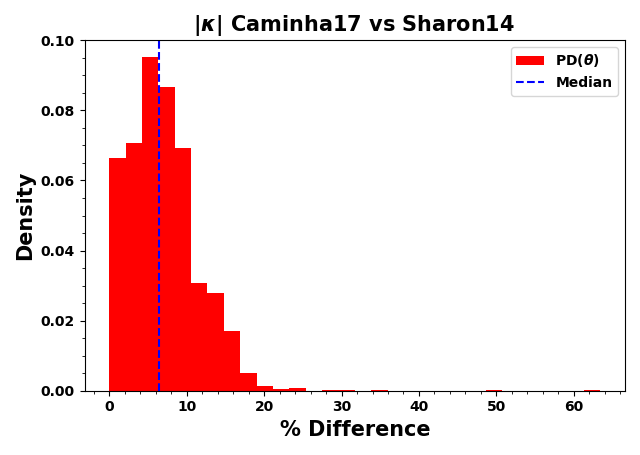}
\caption{Same as Figure \ref{fig:PDhist}, but for the models Sharon14 \citep{johnson14} and Caminha17 \citep{caminha17}. This is an example comparison between two parametric models, which is one of the 3 comparison cases we consider. The MPD between these two models is $6.4\%$ }
\label{fig:ParvsPar}
\end{figure}

\begin{figure}
\includegraphics[trim={1.5cm 0.35cm 1.5cm 0.35cm},clip,width=0.49\textwidth]{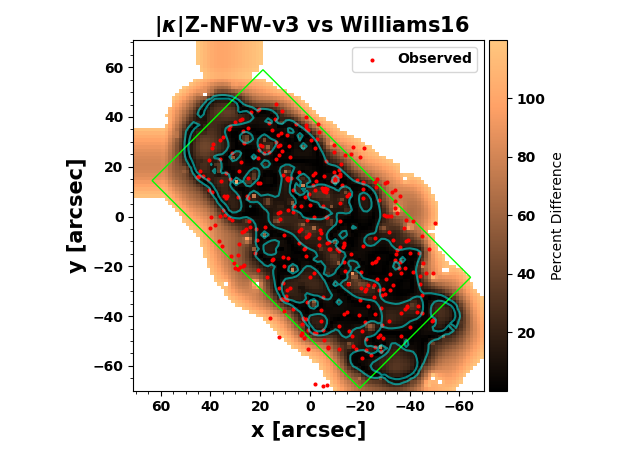}
\includegraphics[trim={0cm 0cm 0cm 0cm},clip,width=0.49\textwidth]{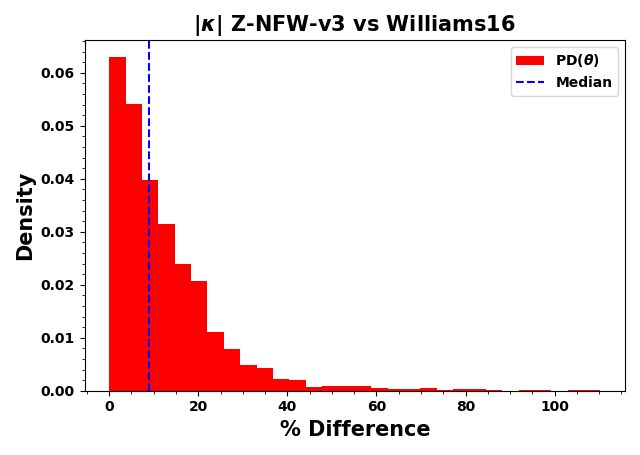}
\caption{Same as Figure \ref{fig:PDhist}, but for the models Z-NFW-v3 \citep{zitrin13} and Williams16 \citep{sebesta16}. This is an example comparison between a parametric and free-form ({\tt GRALE}) model, which is one of the 3 comparison cases we consider. The MPD between these two models is $8.9\%$ }
\label{fig:FFvsPar}
\end{figure}

\begin{figure}
\centering
\includegraphics[trim={3cm 0cm 3cm 0cm},clip,width=0.49\textwidth]{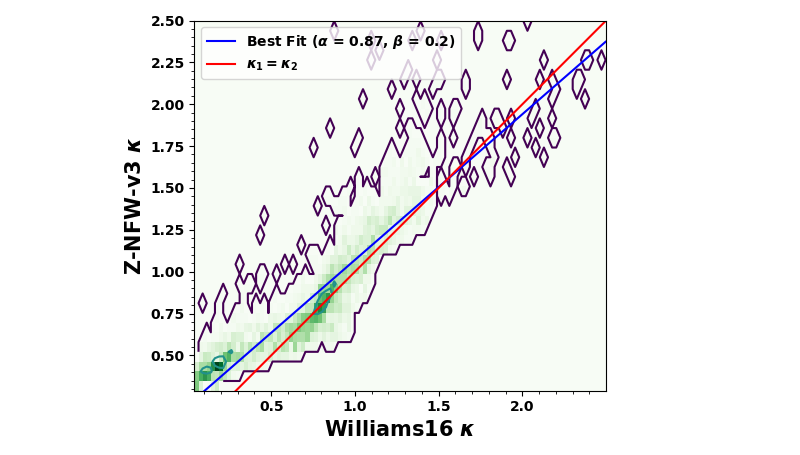}
\includegraphics[trim={3cm 0cm 3cm 0cm},clip,width=0.49\textwidth]{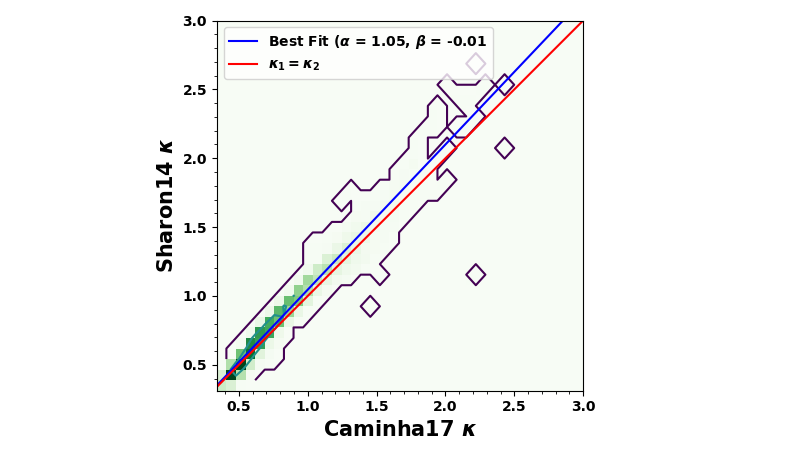}
\caption{Same as Figure \ref{fig:FD} but for Z-NFW-v3 and Williams16 ({\it Left}) and Sharon14 and Caminha17 ({\it Right}). For these comparisons, $\delta_F = 0.200$ for Z-NFW-v3 and Williams16 and $\delta_F = 0.154$ for Sharon14 and Caminha17.}
\label{fig:FDcomparisons}
\end{figure}

Figures \ref{fig:PDhist} and \ref{fig:FD} show necessary plots generated to help visualize the MPD and $\delta_F$ metrics. Those two figures were examples from the comparison between Perera25 and MARS23, a comparison between two free-form reconstruction methods. Here, we provide example MPD and $\delta_F$ plots for comparisons between two parametric methods and free-form and parametric methods. These two cases represent the other main comparison cases we consider in Sections \ref{sec:param} and \ref{sec:paramgrale}. Figure \ref{fig:ParvsPar} shows the MPD comparison plot between Sharon14 and Caminha17, representing the equivalent of Figure \ref{fig:PDhist} for two parametric methods. Figure \ref{fig:FFvsPar} shows the MPD comparison plot between Z-NFW-v3 and Williams16, representing the equivalent of Figure \ref{fig:PDhist} for a parametric and free-form method. Figure \ref{fig:FDcomparisons} shows the corresponding $\delta_F$ plots for the two additional example comparisons, representing the equivalents of Figure \ref{fig:FD}.

\label{lastpage}
\end{document}